\documentclass[aps,prl,twocolumn,showpacs,superscriptaddress]{revtex4}

\def\beq{\begin{equation}}
\def\eeq{\end{equation}}
\def\bea{\begin{eqnarray}}
\def\eea{\end{eqnarray}}
\def\nn{\nonumber}
\def\sss{\scriptscriptstyle}
\def\roughly#1{\mathrel{\raise.3ex\hbox
{$#1$\kern-.75em\lower1ex\hbox{$\sim$}}}}

\def\bd{B_d^0}
\def\bs{B_s}
\def\bdbar{{\bar B}_d^0}

\def\btos{{\bar b} \to {\bar s}}

\def\kbar{{\bar K}^0}

\def\bdKK{{\bd}\to K^0 \kbar}
\def\bdKKstar{{\bd}\to K^{*0} {\bar K}^{*0}}
\def\bsKK{{\bs}\to K^0 \kbar}
\def\Puc{{\cal P}_{uc}}
\def\Ptc{{\cal P}_{tc}}
\def\ZI{Z_{\sss I}}
\def\ZR{Z_{\sss R}}

\begin{document}

\preprint{UdeM-GPP-TH-06-154}
\preprint{UAB-FT-xxx}
\preprint{UMISS-HEP-2006-018}

\title{\boldmath Extracting $\alpha$ from $\bdKK$ Decays}

\author{Alakabha Datta}
\affiliation{Dept of Physics and Astronomy, 108 Lewis Hall, University
of Mississippi, Oxford, MS 38677-1848, USA}
\author{Maxime Imbeault}
\affiliation{Physique des Particules, Universit\'e de Montr\'eal, \\
C.P. 6128, succ.~centre-ville, Montr\'eal, QC, Canada H3C 3J7}
\author{David London}
\affiliation{Physique des Particules, Universit\'e de Montr\'eal, \\
C.P. 6128, succ.~centre-ville, Montr\'eal, QC, Canada H3C 3J7}
\author{Joaquim Matias}
\affiliation{IFAE, Universitat Aut\`onoma de Barcelona, 08193
Bellaterra, Barcelona, Spain}

\date{\today}

\begin{abstract} We propose a new method for obtaining the CP phase
$\alpha$, based on measurements of $\bdKK$, along with theoretical
input.  Due to the similarities of QCD factorization (QCDf) and
perturbative QCD (pQCD), this input is basically the same for each of
these models.  Although the theoretical error is large at present,
many of the contributing quantities will be better known in the
future, leading to a smaller error. One outstanding question is the
extent to which the precision on quark masses, especially $m_c/m_b$,
can be improved.  If one assumes that new physics is not present, the
method can be used to predict the values of CP asymmetries in $\bdKK$:
$0.02 \le A_{dir}^2 + A_{mix}^2 \le 0.125$.  A result outside this
range would signal the existence of long-distance effects beyond those
included in models of nonleptonic decays based on factorization (such
as QCDf and pQCD), or the presence of new physics.
\end{abstract}

\maketitle

A great deal of work has gone into finding a large variety of ways of
measuring $\alpha$, $\beta$ and $\gamma$, the three interior
CP-violating angles of the unitarity triangle \cite{pdg}. The idea is
to find a discrepancy between different ways of measuring the same CP
phase, or with the predictions of the standard model (SM). This would
indicate the presence of physics beyond the SM; it would then be
necessary to include extra (measurable) new-physics parameters in the
amplitude of any process to be analyzed \cite{lmv}.

In the present paper, we present a new method of measuring
$\alpha$. It involves the decay $\bdKK$, along with theoretical
input. At present, the theoretical error is large. However, it is a
function of quantities such as the renormalization scale, light-cone
distributions, and form factors, all of which will become better known
with time. It is also a function of quark masses. If these can be
determined more precisely, especially $m_c/m_b$, this method can
eventually provide a rather accurate way of extracting $\alpha$, and
may help to reveal new physics.

There are several amplitudes contributing to the decay $\bdKK$, which
at the quark level is ${\bar b} \to {\bar d} s {\bar s}$. In
diagrammatic language \cite{GHLR}, the largest is the penguin diagram
${\scriptstyle PEN}$. This amplitude receives contributions from each
of the internal quarks $u$, $c$ and $t$. However, using the unitarity
of the Cabibbo-Kobayashi-Maskawa (CKM) matrix, one can eliminate the
$c$-quark contribution and write
\beq
{\scriptstyle PEN} = V_{ub}^* V_{ud} (P_u - P_c) + V_{tb}^* V_{td}
(P_t - P_c) ~.
\eeq
Note that $P_u$ and $P_c$ arise mainly due to rescattering of the tree
diagrams. The full amplitude for $\bdKK$ can therefore be written
\bea
{\cal A} \equiv {\cal A}(\bdKK) & \!\!=\!\! & V_{ub}^* V_{ud} [(P_u -
P_c) + ...]  \nn\\
& & ~~ + V_{tb}^* V_{td} [(P_t - P_c) + ... ] ~.
\label{bdkkamp}
\eea
In the Wolfenstein approximation \cite{Wolfenstein}, only the CKM
matrix elements $V_{ub}$ and $V_{td}$ are complex.  $V_{ub}^*$ and
$V_{td}$ contain the weak phases $\gamma$ and $-\beta$, respectively,
so that the relative weak phase is $\beta + \gamma = \pi -
\alpha$. The penguin amplitudes $P_i$ contain strong phases. The
amplitude ${\bar A}$ describing the conjugate decay $\bdbar \to K^0
\kbar$ can be obtained from the above by changing the signs of the
weak phases.

The amplitudes ${\cal A}$ and $\bar{\cal A}$ thus depend on four
unknown parameters: the magnitudes $\Ptc\equiv |[(P_t - P_c) + ... ]
V_{tb}^* V_{td}|$ and $\Puc\equiv |[(P_u - P_c) + ...] V_{ub}^*
V_{ud}|$, the relative strong phase $\Delta\equiv {\delta}_{uc} -
{\delta}_{tc}$, and $\alpha$. However, there are only three
measurements which can be made of $\bdKK$: the branching ratio, and
the direct and mixing-induced CP-violating asymmetries.  These yield
the three observables
\bea
X & \!\! \equiv \!\! & \frac{1}{2} \left( |A|^2 + |\bar{A}|^2 \right) \!=\!
\Puc^2 + \Ptc^2 - 2 \Puc \Ptc \cos\Delta \cos\alpha ~, \nn \\
Y & \!\! \equiv \!\! & \frac{1}{2} \left( |A|^2 - |\bar{A}|^2 \right) \!=\!
- 2 \Puc \Ptc \sin\Delta \sin\alpha ~, \\
\label{BKKobservables}
\ZI & \!\! \equiv \!\! & {\rm Im}\left( e^{-2i \beta} A^* {\bar A} \right)
\!=\! \Puc^2 \sin 2\alpha - 2 \Puc \Ptc \cos\Delta \sin\alpha ~. \nn
\eea
In the above, $2\beta$ is the phase of $\bd$--$\bdbar$ mixing.

It is useful to define a fourth observable:
\bea
\ZR & \equiv & {\rm Re}\left( e^{-2i \beta} A^* \bar{A} \right) \\
\label{ZR}
& = & \Puc^2 \cos 2\alpha +\Ptc^2 - 2 \Puc \Ptc \cos\Delta \cos\alpha ~.
\nn
\eea
The quantity $\ZR$ is not independent of the other three observables:
\beq
\ZR^2 = X^2 - Y^2 - \ZI^2 ~.
\eeq
Thus, one can obtain $\ZR$ from measurements of $X$, $Y$ and $\ZI$, up
to a sign ambiguity.

Note that the three independent observables also depend on four
theoretical parameters ($\Puc$, $\Ptc$, $\Delta$, $\alpha$), so that
one cannot obtain CP phase information from these measurements
\cite{LSS}. However, one can partially solve the equations to obtain
\beq
\Ptc^2 = {\ZR \cos 2\alpha + \ZI \sin 2\alpha - X \over \cos 2\alpha -
  1} ~.
\label{Ptceqn}
\eeq

In Ref.~\cite{DL}, two of us (AD,DL) noted that
another constraint on $\Ptc$ can be obtained from $\bsKK$. The
argument goes as follows. To a good approximation, the amplitude for
$\bsKK$ is given by
\bea
{\cal A}(\bsKK) & = & V_{ub}^* V_{us} (P'_u - P'_c) \nn\\
& & ~~~+ V_{tb}^* V_{ts} (P'_t - P'_c) ~,
\eea
where the prime indicates a $\btos$ transition. The first term is much
smaller than the second since $V_{ub}^* V_{us} \ll V_{tb}^* V_{ts}$,
and can be neglected. The decay $\bsKK$ thus basically involves only
$P'_{tc} \equiv |(P'_t - P'_c) V_{tb}^* V_{ts}|$, and this quantity
can be obtained from its branching ratio. In the limit of perfect
flavour SU(3) symmetry, $P'_{tc} = \Ptc$, apart from known CKM matrix
elements. Thus, the measurement of $P'_{tc} $ gives the necessary
additional condition for extracting $\alpha$ from $\bdKK$ (a
connection with $B_d \to \pi\pi$ can be found in \cite{fl}). However,
SU(3)-breaking effects will affect the relationship between $P'_{tc}$
and $\Ptc$. These are $O(30\%)$ and are unknown, leading to a large
theoretical error. In Ref.~\cite{DL} this error was reduced to
$O(5\%)$ by considering two decays, e.g.\ $\bdKK$ and $\bdKKstar$, and
using double ratios.

In the present paper, we point out that there is an alternative
additional piece of information. In order to see this, it is simpler
to eliminate the $t$-quark contribution from ${\scriptstyle PEN}$,
writing
\beq
{\cal A}(\bdKK) = V_{ub}^* V_{ud} T + V_{cb}^* V_{cd} P ~,
\eeq
where $P \equiv (P_c - P_t) + ...$ and $T = (P_u - P_t) + ...$ are
complex quantities. (The $...$ represent smaller contributions.)
Compared to the previous parametrization [Eq.~(\ref{bdkkamp})], we see
that $|P \, V_{tb}^* V_{td}| = \Ptc$, while $|(T - P) V_{ub}^* V_{ud}|
= \Puc$.

As we will see below, the theoretical input is a prediction of the
quantity $|T-P|$. There are basically two competing models for
calculating nonleptonic decays: QCD factorization (QCDf)
\cite{BBNS} and perturbative QCD (pQCD) \cite{pQCD}. An
effective-theory approach to nonleptonic decays that incorporates
the ideas of QCDf and pQCD is the soft collinear effective theory
(SCET) \cite{SCET}. Here, we will only consider QCDf and pQCD.

Many references have concentrated on the differences between QCDf and
pQCD. However, here we focus on their similarities. For example, both
are models of nonleptonic $B$ decays where the long-distance and
short-distance physics are factorized at the $b$ scale. The
short-distance physics is calculable as an expansion in the strong
coupling constant, while the long-distance effects are encoded in the
form factors, light-cone distributions, etc. In particular, both QCDf
and pQCD have the same mathematical formula for $T-P$. As such,
despite the different methods of calculation, the prediction of the
central value and error on $|T-P|$ will be quite similar in QCDf and
pQCD. We treat these models in turn in what follows.

We begin with QCD factorization. QCDf finds that
\bea
{T \over {A^0_d}} & \!\!=\!\!& \alpha_4^u - \frac{1}{2}\alpha_{4EW}^u +
\beta_3^u + 2 \beta_4^u - \frac{1}{2} \beta^u_{3EW} - \beta^u_{4EW} ~,
\\
{P \over {A^0_d}} & \!\!=\!\!& \alpha_4^c - \frac{1}{2}\alpha_{4EW}^c +
\beta_3^c + 2 \beta_4^c - \frac{1}{2} \beta^c_{3EW} - \beta^c_{4EW} ~,
\eea
where the normalization factor in the denominator is $A^0_d=M^2_{\bd}
F_0^{\bd\to \kbar}(0) f_K {G_F}/{\sqrt{2}}$. The quantities
contributing to $T / {A^0_d}$ and $P / {A^0_d}$ are defined and
computed in Refs.~\cite{BBNS,BN}.

In QCDf the various hadronic quantities are calculated using a
systematic expansion in $1/m_b$. However, a potential problem occurs
because the higher-order power-suppressed hadronic effects contain
some chirally-enhanced infrared (IR) divergences. In order to
calculate these, one introduces an arbitrary IR cutoff. The key
observation here is that the difference $\Delta_d \equiv T - P$ is
free of these dangerous IR divergences. This difference has been
calculated \cite{DMV}.  The latest value of $|\Delta_d| \equiv |T -
P|$, which is the only physical quantity, is:
\bea
|\Delta_d| & = & (2.96\pm 0.97) \times 10^{-7} ~{\rm GeV} ~.
\label{Deltad}
\eea
Note that in the difference $T-P$, the effect of electroweak
contributions has been neglected in this calculation, but their impact
is expected to be very tiny.

The fact that $|\Delta_d|$ is known and is free of IR divergences
was first used in Ref.~\cite{DMV} to predict the values of the
various quantities in $\bs \to K {\bar K}$. Note that in a
$\Lambda/m_b$ expansion, we can expect corrections of up to of
$O(10)\%$ with respect to the leading term for a generous range of
values of $\Lambda$ and $m_b$, although they are often found to be
smaller than expected by dimensional arguments (for instance, see
Ref.~\cite{ball}). However, this is not true if power-suppressed
IR divergences arise. These can induce errors up to 100\% (for
example, see Ref.~\cite{BN}). Although $\Delta_d$ has a residual
sensitivity to the finite and suppressed $\Lambda/m_b$
corrections, being free from  IR divergences makes it more robust.
Other ways of trying to approach the problem of $1/m_b$
corrections can be found in Ref.~\cite{hf}.

In pQCD, the calculation of various hadronic quantities is different
than in QCDf. In particular, all quantities are finite. Thus, the
calculation of $T$ and $P$ in pQCD is different than in QCDf. However,
the difference $T-P$ is the same in both formulations. Indeed, in
Ref.~\cite{LMS}, it is noted that the mathematical expression for
$T-P$ is the same as in QCDf. In calculating this expression, the only
difference is the renormalization scale, $\mu$. QCDf takes $ \mu \sim
m_b$~\cite{co}, while pQCD
uses $\mu^2 = \Lambda_{\sss QCD} m_b$.  Although the
difference $T-P$ has not been computed explicitly in pQCD, the above
suggests that its value is similar to that of QCDf.  For this reason,
in what follows we use the QCDf calculation and assume that
$|\Delta_d|$ is as in Eq.~(\ref{Deltad}).

At present, the errors on the quantities in $|\Delta_d|$ are quite
large so that the extraction of $\alpha$ (below) suffers significant
theoretical uncertainty. There are four sources of error. They are (i)
the renormalization scale $\mu$, which enters $\alpha_s(\mu)$,
$C_1(\mu)$ \cite{BBNS} and $r_\chi^{K}(\mu)$ \cite{BN}, (ii) the
Gegenbauer coefficients $\alpha_1^{K}$ and $\alpha_2^{K}$ which enter
the light-cone distributions (their dependence on $\mu$ is irrelevant
here given the large error taken for them), (iii) the form factor
$F_0^{B\to K}$ \cite{BN}, and (iv) the quark masses ${\bar m}_s(2~{\rm
GeV})$ \cite{pdg} and $m_c/m_b$. Note that the same sources of error
also affect the pQCD calculation. In calculating the errors in
$|\Delta_d|$, these quantities are taken to lie in the following
ranges \cite{pdg,BN}: $m_b/2<\mu<2 m_b$, $\alpha_1^K=0.2 \pm 0.2$,
$\alpha_2^K=0.1\pm 0.3$, $F_0^{B\to K}=0.34\pm0.05$, ${\bar
m_s}(2~{\rm GeV})=103 \pm 20$ MeV, $0.26<m_c/m_b<0.36$.

It should be noted that in pQCD, the form factors are calculable in
terms of the light-cone distributions of the mesons \cite{sanda}. In
this approach it is assumed that the soft contributions to the form
factors are suppressed.  There is no rigorous estimate of the error in
the form-factor predictions from this assumption.  In the QCDf
approach the form factors are inputs which may be taken from QCD sum
rule calculations \cite{ball}. The sum rule approach assumes that the
soft contributions dominate the form-factor calculation. The inputs to
this calculation then include, in addition to the light-cone
distributions, additional parameters associated with the Borel
transformation. Here too there is no rigorous estimate of the error in
the form-factor calculations from the assumptions involved in this
approach. The predictions of the form factors from the sum rule and
the pQCD approach are similar and it is reasonable to guess that the
error in the form-factor calculation in both approaches is similar.
Rigorous information about the form factors may be obtained from
lattice calculations in the future with further progress in this
area. Alternately, the form factors may be measured in experiments
where one can have access to $ B \to K$ form factors in $ B \to K ll$
decays.  One can also relate the $B \to K$ form factors to $ B\to \pi$
form factors, obtained from semileptonic decays, by using SU(3)
symmetry if the SU(3)-breaking effects can be reliably estimated.

\begin{table}[!tbp]
\begin{center}
{
\begin{tabular}{|l|c|c|c|c|c|c|}
\hline
                 & $\mu$ &
                        $\alpha_1^K$
                        &  $\alpha_2^K$ &
                        $F_0^{B\to K}$     &
                        ${\bar m_s}(2~{\rm GeV})$    &
                        $m_c/m_b$        \\
\hline $|(\Delta_d)|$     & $19.7\%$ &
                         $4.1\%$, & $0.9\%$& $15.8\%$ & $5.7\%$ &
$53.7\%$  \\
\hline
\end{tabular}
}
\end{center}
\caption{Relative impact of each error on $|\Delta_d|$, defined as
$\sigma^2_i/\sigma^2_{total}$, with $i = \mu$, $\alpha_1^K$,
$\alpha_2^K$, $F_0^{B\to K}$, ${\bar m_s}$, $m_c/m_b$.}
\end{table}

Table 1 indicates the relative error associated with each of the
inputs entering $|\Delta_d|$.  The error related to $\mu$ is quite
large. However, this error will be reduced when higher-order
calculations are done.  Concerning the Gegenbauer coefficients, to be
conservative we prefer to use the values given in Ref.~\cite{BN} even
if the errors are a bit too large. There have been several interesting
developments on the lattice \cite{lat} and with QCD sum rules
\cite{khodball}, leading to more precise numbers for these
coefficients.  However, the problem is always how to estimate the
error associated with the higher Gegenbauer modes not included in the
truncated series. Hopefully, this difficulty will be resolved in the
future.  The largest source of error is the ratio $m_c/m_b$. Clearly a
greater control of $m_c$ will lead to a substantial reduction of this
error.  A better understanding of nonperturbative effects may help in
the extraction of $m_c$ from charmonium data \cite{charm}.

Now, both QCDf and pQCD claim that, at leading order in $1/m_b$, the
short- and long-distance physics in nonleptonic decays factorize. If
nonfactorizable long-distance effects are important, the predicted
value of $|\Delta_d|$ will change. Thus, any method of determining
weak-phase information that uses input from theoretical models based
on factorization must also provide an independent test of such models
which does not use that phase information. As we will see, the method
described in this paper does just that. Thus, we propose a new way of
obtaining $\alpha$ {\it and} a clean test of theoretical hadronic
models based on factorization.

We have shown that $\Ptc$ can be defined in terms of $\alpha$ and the
experimental $\bdKK$ observables [Eq.~(\ref{Ptceqn})]. $\Puc$ can be
obtained similarly:
\beq
\Puc^2 = {\ZR  - X \over \cos 2\alpha -  1} ~.
\label{alphaeqn}
\eeq
Now, using the relation $|(T - P) V_{ub}^* V_{ud}| = \Puc$, one can
obtain the weak phase $\alpha$.

In order to extract $\alpha$, one proceeds as follows. It is first
necessary to relate the quantities $X$, $Y$, $\ZI$, $\ZR$ to the
measurements. It was noted in Ref.~\cite{CGRS} that
\bea
\Gamma (B \to f) &=& \frac{p_c}{8 \pi m_B^2} |A(B \to f)|^2 ~, \nn\\
BR(B \to f) &=& \Gamma (B \to f) \frac{\tau_0}{\hbar} ~.
\eea
where $p_c$ is the momentum of the final-state mesons in the rest
frame of the $B$, and $m_B$ and $\tau_0$ are the mass and lifetime of
the $B^0$ meson, respectively.  A similar expression holds for the
averaged branching ratio, which includes both $A$ and ${\bar A}$.
Thus, we have
\bea
X = \frac{8 \pi m_B^2 \hbar}{\tau_0 p_c} \mathcal{B} \equiv \kappa
\mathcal{B}~,
\eea
where $\mathcal{B}$ is the averaged branching ratio.  For $Y$, we have
\beq
\frac{Y}{X} = A_{dir} \Rightarrow Y = \kappa \mathcal{B} A_{dir}~.
\eeq
Now recall that $A_{mix} = - 2 \, {\rm Im} (\lambda) / ( {1 +
  |\lambda|^2} )$, where $\lambda = e^{-2 i \beta} \bar A / A$. Thus,
\beq
\ZI = |A|^2 {\rm Im}{\lambda} = - X A_{mix} =- \kappa \mathcal{B} A_{mix}~.
\eeq
Combining these expressions for $X$, $Y$ and $\ZI$, we get
\beq
\ZR = \pm \kappa \mathcal{B} \sqrt{1 - A_{dir}^2 - A_{mix}^2}~.
\eeq
Certain references adopt a different convention for $A_{mix}$ which
differs from that given above by a sign. However, as can be seen from
the above expression for $\ZR$, this sign is unimportant here.
Eq.~(\ref{alphaeqn}) can now be solved for $\alpha$:
\bea
1- \cos 2\alpha & = & 2 \sin^2 \alpha \nn\\
& & \hskip-1.5truecm = \frac{\kappa \mathcal{B}}{|V_{ub}^* V_{ud}
  |^2|\Delta_d|^2} \left( 1 \pm \sqrt{1 - A_{dir}^2 - A_{mix}^2}
\right).
\label{reln}
\eea
Note that Ref.~\cite{DMV} also shows that there exists a relation
(called a sum-rule) between $|\Delta_d|$, the branching ratio and
CP asymmetries of $\bdKK$, and $\gamma$. The above equation is
equivalent to this sum-rule. (For a similar relation for the weak
mixing angle $\phi_s$ see \cite{phis}).

The various experimental quantities in $\bdKK$ have now been measured
\cite{BdKKBR}. They are
\bea
\mathcal{B} & = & 0.95^{+0.20}_{-0.19} \times 10^{-6} ~, \nn\\
A_{dir} & = & 0.58^{+.66}_{-.73} \pm 0.01 ~[Belle], \nn\\
A_{dir} & = & -0.40 \pm 0.41 \pm 0.06 ~[Babar], \nn \\
A_{mix} & = & 1.28^{+0.73}_{-0.80}{}^{+0.16}_{-0.11} ~.
\eea
Combining these measurements (so far not very precise) with the value
of $|V_{ub}^* V_{ud}| = (3.586^{+0.104}_{-0.076}) \times 10^{-3}$
\cite{CKMfitter}, and that for $|\Delta_d|$ given in
Eq.~(\ref{Deltad}), we can obtain $\sin^2 \alpha$. Unfortunately, at
present, there is no constraint; all values $0 \le \sin^2 \alpha \le
1$ are allowed.

There is, however, more that can be learned. If one assumes that there
is no new physics, then one can use other determinations of the weak
phases. Along with the measurement of the $\bdKK$ branching ratio and
the theoretical input of $|\Delta_d|$, knowledge of the weak phases
allows one to make predictions for the $\bdKK$ CP asymmetries
[Eq.~(\ref{reln})]. At present, $\beta$ and $\gamma$ are measured more
accurately than $\alpha$, so it is useful to write $\sin^2 \alpha =
\sin^2(\beta+\gamma)$. Taking (in degrees) $\beta =
22.03^{+0.72}_{-0.62}$ and $\gamma = 59.0^{+9.2}_{-3.7}$
\cite{CKMfitter}, the SM currently gives $0.95 \le \sin^2 \alpha \le
1$. Now, the value of the coefficient $(\kappa \mathcal{B})/(|V_{ub}^*
V_{ud} |^2|\Delta_d|^2) $ is $98 \pm 67$.  This requires $A_{dir}$ and
$A_{mix}$ to be small, with the negative sign of the square root, in
order for the right-hand side to reproduce the SM value of $\sin^2
\alpha$. In particular, from Eq.~(\ref{reln}) we obtain $0.02 \le
A_{dir}^2 + A_{mix}^2 \le 0.125$. This then is a {\it prediction} of
this method. Put another way, this is a {\it test} of theoretical
hadronic models based on factorization.  If this constraint on
$A_{dir}^2 + A_{mix}^2$ is not satisfied, it means that the expected
SM value of $\sin^2 \alpha$ is not reproduced, which in turn implies
that the value of $|\Delta_d|$ [Eq.~(\ref{Deltad})] is wrong (or there
is new physics).

If the test fails, it means there are effects that QCDf and pQCD have
not considered, i.e.\ nonfactorizable long-distance physics. One
example of such an effect could be possible long-distance
contributions from $c \bar{c}$ intermediate states to the $c$-quark
piece of the penguin diagram, $P_c$ (``charming penguins''
\cite{charming, SCET}). In this case, the calculation of $|\Delta_d|$
in Eq.~(\ref{Deltad}) is not correct, as it does not take these
effects into account, and the test will fail. Though the impact of
charming penguins on $B$ decays is a controversial issue
\cite{contro}, the method proposed in this paper for extracting
$\alpha$ can be understood also as a test to find long-distance
effects such as charming penguins.

Note that if long-distance effects turn out to be important, the
method proposed here for determining $\alpha$ can still be used if a
reliable theoretical evaluation of $T-P$ which includes the
long-distance effects can be found.

In summary, in this paper we have proposed a new method of obtaining
the CP phase $\alpha$. It involves measurements of the decay $\bdKK$,
along with theoretical input. This input is largely the same whether
it is taken from QCD factorization (QCDf) or perturbative QCD (pQCD).
The theoretical error is very large at present. We have computed the
percent errors due to the various contributions. We note that the
theoretical error depends on quantities such as the renormalization
scale, light-cone distributions, and form factors, which will all be
better known in the future, so that this error will be correspondingly
reduced.  Unfortunately, the theoretical error also depends strongly
on quark masses, particularly $m_c/m_b$. It appears difficult to
increase the precision on this ratio, but if this can be done this
method can eventually provide a new way of extracting $\alpha$, and
may help to reveal the presence of new physics.

If one assumes that new physics is not present, the method allows one
to make predictions for the CP asymmetries in $\bdKK$ decays. This
provides a test of the standard model, and permits one to distinguish
between factorization models of nonleptonic $B$ decays (such as QCDf
and pQCD) and those which include long-distance effects. In this way
it may be possible to ascertain which kind of physics is present in
$b$ decays.

\medskip
\noindent {\bf Acknowledgements}:
This work was financially supported by NSERC of Canada (MI \& DL),
and by FPA2005-02211, PNL2005-51 and the Ramon y Cajal Program
(JM).
\\



\begin{thebibliography}{References:}
\bibitem{co} Note that QCDf annihilation and hard-scattering uses a
different scale, $\mu^2 = \Lambda_{\sss QCD} m_b$.
\end{thebibliography}
\end{document}